# Influence of generated defects by Ar-implantation on the thermoelectric properties of ScN

Razvan Burcea[1], Jean-François Barbot[1, *], Pierre-Olivier Renault[1], Dominique Eyidi[1], Thierry Girardeau[1], Marc Marteau[1], Fabien Giovannelli[2], [3]Ahmad Zenji[3], Jean-Michel Rampnoux[3], Stefan Dilhaire[3], Per Eklund[4] and Arnaud Le Febvrier[4, *]

[1] *Institute PPRIME, CNRS, Université de Poitiers–ENSMA, UPR 3346, SP2MI, TSA 41123, 86073 Poitiers cedex 9, France*
[2] *Laboratoire GREMAN, CNRS, Université de Tours, UMR 7347, 41029 Blois Cedex, France*
[3] *Laboratoire LOMA, CNRS, Université de Bordeaux, UMR 5798, 33405 Talence, France*
[4] *Department of Physics, Chemistry and Biology (IFM), Linköping University, SE-581 83 Linköping, Sweden*

## Abstract:

Nowadays, making thermoelectric materials more efficient in energy conversion is still a challenge. In this work, to reduce the thermal conductivity and thus improve the overall thermoelectric performances, point and extended defects were generated in epitaxial 111-ScN thin films by implantation using argon ions. The films were investigated by structural, optical, electrical, and thermoelectric characterization methods. The results demonstrated that argon implantation leads to the formation of stable defects (up to 750 K operating temperature). These were identified as interstitial type defect clusters and so-called argon-vacancy complexes. The insertion of those specific defects induces acceptor-type deep levels in the bandgap yielding to a reduce of the free carrier mobility. With a reduced electrical conductivity, the irradiated sample exhibited higher Seebeck coefficient while maintaining the power factor of the film. The thermal conductivity is strongly reduced from 12 to 3 $W.m^{-1}.K^{-1}$ at 300 K, showing the influence of defects in increasing phonon scattering. Subsequent high temperature annealing, at 1573 K, leads to the progressive evolution of those defects: the initial clusters of interstitials evolved to the benefit of smaller clusters and the formation of bubbles. Thus, the number of free carriers, the resistivity and the Seebeck coefficient are almost restored but the mobility of the carriers remains low and a 30% drop in thermal conductivity is still effective ($k_{total}$ ~ 8.5 $W.m^{-1}.K^{-1}$). This study shows that the control defect engineering with defects introduced by irradiation using noble gases in a thermoelectric coating can be an attractive method to enhance the figure of merit of thermoelectric materials.

*Corresponding author:

arnaud.le.febvrier@liu.se

jean.francois.barbot@univ-poitiers.fr

# 1 - INTRODUCTION

Defects, inevitable in materials, are classified as point defects (vacancy, interstitial, impurity), linear (dislocation), planar (stacking fault, grain boundary) and volumetric defects (from nanocavity up to microcrack). They are known to influence properties of materials by themselves and by interacting with each other. Some specific defects can also be introduced deliberately, particularly after the material has been subjected to irradiation/implantation or plastic deformation. For use, impurities are introduced into semiconductors to tailor their electrical conductivity $\sigma$. On the contrary in metals such impurities can reduce the conductivity by acting as obstacles for a smooth carrier flow and play a key role in extrinsic phonon scattering in lowering the thermal conductivity $k_{total}$. The ion implantation technique in which specific elements can be selected and then accelerated toward a target material is widely used in several fields as the microelectronic industry, the surface modification or to simulate the behavior of materials under harsh environment (nuclear applications). The process results in the generation of a large concentration of Frenkel pairs via the collision cascades, which tend to recombine and/or condense to form various type of defects depending on the experimental conditions such as the species of implanted atoms, the fluence, the incident energy and the thermal budget. An advantage of ion implantation is that the as-induced changes in physical properties can be manipulated repeatedly which makes it a suitable technique for dealing with the transport properties of thin film thermoelectric (TE) materials. Indeed, the efficiency of a TE-material to convert heat to electricity is governed by its figure of merit $ZT = S^2\sigma T/k_{total}$, where $T$ is the absolute operating temperature, $S$ the Seebeck coefficient, $\sigma$ the electrical conductivity and $k_{total}$ the thermal conductivity being the sum of the lattice and electronic thermal conductivity ($k_{lat} + k_e$). The nanoscale defects generated by implantation (may be followed by high temperature annealing) could be used as a scattering means for short wavelength phonons leading to a reduction of $k_{lat}$. Besides, defects are currently seen as a promising way forward in a strategy to improve the efficiency of TE-materials.[1] Vacancy engineering strategy has been used to create dislocations in PbTe for reducing the lattice thermal conductivity.[2] Similarly, introduction of excess Cu atoms in CuSe thin films reduces the Cu-vacancies and introduces additional scattering centers, increasing the $ZT$.[3] It should be noted that other strategies for enhancing $ZT$ have been developed as for example the quantum confinement,[4,5] the use of antisite defects in ZrNiSn half-Heusler alloys or incorporation nano-inclusions in the TE matrix.[6,7] The controlled introduction of nanoscale defects by using ion implantation can also be part of the strategies to scatter phonons reducing the thermal conductivity in thin film TE-materials.

This ion beam technique has recently been used in TE-films to modify the electronic transport properties in relation with the microstructural modifications. For example, the $S^+$ implantation in $Bi_{0.5}Sb_{1.5}Te_3$ increases the thermoelectric power factor, $PF = S^2\sigma$, via the generation of carriers due to the $Te_{Bi}$ antisites.[8] Similarly, the N implantation in $SrTiO_3$ creates mainly oxygen vacancies acting as carrier donors while decreasing the grain size leading to an enhancement of PF value.[9] In $CoSb_3$ the $Fe^+$ implantation changes the conductivity type due to the creation of vacancies and an increase of PF by more than a factor of 10 is reported.[10] All these authors highlight the defects engineering in the applications of TE-devices by an increase of the PF. It should be noted, however, that it is mainly the direct or indirect doping effect that is at the origin of the modifications of the electronic properties. These are strongly dependent on the implantation conditions, and in particular on the dose. Ion implantation can also lead to the formation of distinct phases, as the $Ag_2Te$ phases in Ag-implanted PbTe, resulting in an increase in the Seebeck coefficient.[11] In summary, implantation in TE-films was used as a means of controlling charge carrier properties but not as a means of introducing lattice defects to strength phonon scattering.

Scandium nitride (ScN) is a n-type semiconductor with suitable properties such as a high carrier concentration in the range $10^{18}$-$10^{22}$ cm$^{-3}$ and low electrical resistivity of about 300 µΩ.cm leading to an appreciable PF of about $3\times10^{-3}$ Wm$^{-1}$K$^{-2}$ at 600 K.[12,13] However, due to its high thermal conductivity, the overall $ZT$ is limited, in the range 0.2-0.3.[13-15] The total thermal conductivity, mainly dominated by the lattice conductivity $k_{lat}$,[16] is found to be in the range 10-12 Wm$^{-1}$K$^{-1}$ at room temperature (RT) and decreases with increasing temperature due to Umklapp scattering (7-8 Wm$^{-1}$K$^{-1}$ at 500 K).[15] Attempts were made to reduce the thermal conductivity by alloy scattering such as the introduction of Nb (~10 at. %) which leaded to a large decrease of the thermal conductivity, down to 2.2 Wm$^{-1}$K$^{-1}$ but deteriorated the Seebeck effect.[17] More recently it has been shown that defect introduction by using Mg-dopant implantation leads to an increase in the Seebeck coefficient coupled with a drop in thermal conductivity $k_{total}$, down to 3.2 Wm$^{-1}$K$^{-1}$ for the ScN sample implanted with 2.2 at. % of Mg.[18] Another study has shown the potential on Li$^+$-implanted ScN for which the thermal conductivity is divided by half in the 300-700K temperature range.[19] Defect engineering by ion implantation and other techniques have shown potential for improving of thermoelectric properties. However, the control of the induced defect in a material is always challenge in terms of their formation during irradiation/implantation, the type of defect, and their stability with temperature which is critical for thermoelectric applications.

The underlined idea of the present work is to reduce the lattice thermal conductivity, $k_{lat}$, of ScN by introducing a network of lattice defects (acting as phonons scattering centers) via the ion implantation while trying to keep the power factor constant and thereby improve the *ZT* value. To promote the introduction of defects and minimize the chemical doping effects of the implanted species, a heavy noble gas, argon, was implanted in the ScN thin films. The effects of post implantation annealing up to 1573 K were also investigated to discuss the implantation-induced defects evolution in relation to changes in thermoelectric properties. In many materials, metals and semiconductors,[20,21] the interaction of gas atom with excess vacancies leads to the formation of nanoscale bubbles that can turn into voids if desorption takes place under subsequent high temperature annealing. Results show that these Ar-implantation induced-defects modify the physical properties of ScN films by reducing the thermal conductivity while maintaining roughly constant the power factor, thus showing their potential for improving *ZT*.

## 2 - MATERIAL and METHODS

<111> degenerate n-type ScN thin films (thickness ~ 240nm) were deposited using dc reactive magnetron sputtering onto $Al_2O_3$ (c-cut) substrates maintained at a temperature of 800°C (for more details on the growth conditions see[13]). The thin films were then implanted with argon ions, $Ar^{2+}$, at room temperature using the implanter EATON VN3206 at Pprime Institute (Poitiers). The depth profiles of the displaced atoms and implanted ions in the ScN films (density of 4.29 g/cm$^3$) were calculated using the SRIM 2013 software under the full-damage cascade mode.[22] To introduce a constant quantity of damage (called displacements per atom: dpa) along the thickness of the film, a multi-implantation protocol was designed. Three implantations with decreasing incident energies of 320 keV (projected ion mean range $R_p$ ~ 200 nm with straggling $\Delta R_p$ ~ 60 nm), 160 keV ($R_p$ ~ 100 nm and $\Delta R_p$ ~ 35 nm) and 50 keV ($R_p$ ~ 35 nm and $\Delta R_p$ ~ 15 nm) were carried out with fluences of 3.5x10$^{15}$, 1x10$^{15}$ and 1x10$^{15}$ cm$^{-2}$, respectively. These relevant fluences were chosen to implant the ScN film in the called high damage regime, i.e. 5-6 dpa, for which a previous study showed the impact of defects on thermal conductivity is the most significant.[18] Fig.1 shows the resulting damage distribution (dpa) and argon concentration. As seen, the dpa-profile is rather flat in the entire ScN film, around 5 dpa, while the argon concentration is bumpy in the range $0.1 - 0.3\ at.\%$ and extends deep into the substrate. During implantation, the current beam density was kept below 5 µA.cm$^{-2}$ to avoid any temperature increase. SRIM calculations which do not consider any dynamic recombination, result in a vacancies/ion ratio of 840 (for a given energy of 50 keV) showing that the defect formation is promoted over the effects of the implanted impurity. Subsequent annealing at 1573 K were conducted in a home-made lamp furnace under ambient atmosphere. The rate of heating was about 20°C/min and the annealing duration of 10 min.

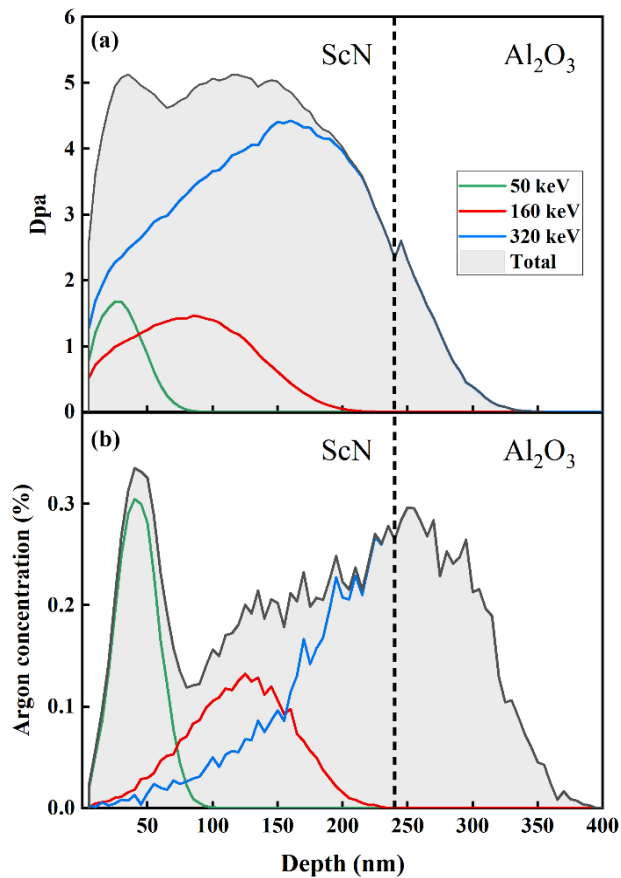

*Figure 1 : (a) Displacement per atom (dpa) and (b) Ar concentration profiles as function of depth calculated by using the SRIM code. The ScN film was implanted with 320-160-50 keV Ar-ions at fluences of 3.5x10$^{15}$, 1x10$^{15}$ and 1x10$^{15}$ cm$^{-2}$ respectively to obtain a flat damage profile of around 5 dpa throughout the film (grey).*

The macroscopic in-plane resistivity $\rho(T)$ and mobility $\mu(T)$ were measured using by the van der Pauw method coupled with Hall effect (ECOPIA HMS-5000). Two cryostats were used: a low temperature cryostat, from 80 to 350 K and a high temperature cryostat, from 300 to 750

K. All these measurements were performed using a constant magnetic field up to 0.580 Tesla. The rate of temperature increase during measurements was close to 3 °C min$^{-1}$.

Optical measurements were carried out by using a J. A. Woollam M2000XI ellipsometer in the range 0.2-1.7µm. Ellipsometric data were acquired at 55°, 65° and 75° angles of incidence. To determine the optical properties, a three-oscillator model was developed using the J. A. Woollam CompleteEase software: a Tauc−Lorentz oscillator (TLO) centered close to 2.4 eV modelling the direct band gap absorption of ScN; a Gaussian oscillator (GO) arbitrarily centered out of the measurement range at 7 eV to model all UV interband transitions; and a Drude oscillator (DO) modelling the free-carrier optical behavior in the NIR range. The optical properties of the $Al_2O_3$ substrate were modelled using the optical constants available in the J. A. Woollam database.

XRD measurements were performed on a four-circle diffractometer (Seifert Space XRD TS-4) with a Cu X-ray source using a 0.5 mm collimator and a Meteor0D detector. The residual stress was analyzed using the sin² Ψ-method, which relies on the use of lattice plane spacing $d_{hkl}$ as an internal strain gauge. Along a given direction ($\Psi, \Phi$), where $\Psi$ is the angle between the surface normal and the normal to (*hkl*) planes and $\Phi$ is the azimuthal angle, the measured lattice strain $\varepsilon_{\Psi,\Phi}$, is given by:

$$\varepsilon_{\psi,\phi} = \frac{a_{\psi,\phi} - a_0}{a_0} \qquad (1)$$

where $a_0$ is the stress-free lattice parameter and $a_{\Psi,\Phi}$ is the lattice parameter determined from a given {*hkl*} reflection. The $a_0$ parameter is generally unknown and might differ significantly from the $a_{bulk}$ parameter, preventing a direct determination of the strain.

Transmission electron microscopy (TEM) data were acquired using a TALOS F200S Thermofisher microscope operating at 200 kV. Slices, with a thickness of about 300 µm, were cut from the bulk sample. They were then pre-thinned up to a thickness of approximately 20 µm by mechanical polishing and glued onto a copper grid. Finally, ion thinning down to electron transparency was performed by means of a Precision Ion Polishing System (Gatan-PIPS). A TEM thin foil was also extracted by focused ion beam (FIB) Helios G3 CX - from Thermofisher Dual Beam. The lamella was cut perpendicular to the basal plane, and was about 12 µm long, 4 µm wide and 80 nm thick (approximate values).

The in-plane Seebeck coefficient and the electrical resistivity were measured simultaneously under a low-pressure helium atmosphere using ULVAC-RIKO ZEM3 from RT up to 680 K.

Thermal conductivity measurements were performed at room temperature by Frequency Domain Thermoreflectance (FDTR) setup. This technique is non-contact and nondestructive optical methods. The measurement of the temperature oscillation induced by the absorption of an intense modulated laser beam (pump) allows its thermal characterization. A thin metallic film (Aluminum, 67 nm) was deposited on top of the sample to confine the heat absorption and to sense the surface temperature by its reflectance change. The modulated CW pump laser is focused on top of the surface of the sample. The absorbed energy induces a periodic temperature change. A second CW laser beam (probe) overlaps the heating area and is reflected to a photodiode. The relative variation of the probe intensity ($\Delta I / I_0$) is detected with a lock-in amplifier synchronized on the pump frequency. The pump frequency ($f$) is swept from a few kHz up to 100 MHz and the best fit of the spectral response is then performed with multilayered model. The thermal conductivity ($k_{total}$), the heat capacity ($C_{TH}$) and the thermal contact resistance ($R_{contact}$) are thus obtained.

## 3 - RESULTS

### 3.1 - STRUCTURAL CHARACTERIZATION

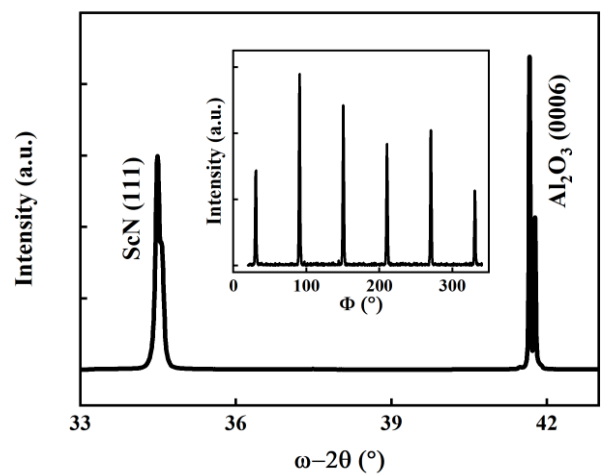

*Figure 2: ω-2θ X-ray diffraction pattern (off set 0.2°) from a ScN (111) film deposited onto an Al₂O₃ (0006) substrate. The inset shows a Φ scan at Ψ =70.5°.*

Figure 2 shows a $\omega$-$2\theta$ (off set 0.2°) XRD scan of the as-deposited ScN thin film on sapphire substrate. The peak observed at 34.5° is identified as (111) ScN with a lattice parameter of 4.50 Å, in good agreement with ICCD PDF 00-045-0978 (ScN). The inset shows the $\phi$ scan at azimuthal angle $\Psi = 70.5°$ for ScN {111}. The presence of six peaks shows that ScN thin film grows epitaxially with the [111] out of plane direction and the presence of twin domains usually observed for the growth of cubic material system on c-plane sapphire.[13,23]

Figure 3 compares the diffraction peaks of the 333-diffraction peak for the as-deposited (reference), implanted and annealed ScN samples. Even present at high 2θ angle, the 333 reflection was chosen to observe more clearly any changes (shift, shapes intensity) compared to other lll reflection. As seen, the reference sample exhibits the two peaks from the non-monochromatic X-ray source ($K_{\alpha 1}$ and $K_{\alpha 2}$) showing the good crystallinity of the film. After implantation, a large drop in intensity is observed and the $K_{\alpha 1}$ and $K_{\alpha 2}$ split is no longer visible, suggesting a drastic change of the structure. The FWHM is also strongly enlarged meaning that the local strain heterogeneities are increased by the implantation. A shift towards the lower $2\theta$ angles is observed indicating that the interplanar distance $d_{333}$ is expended by the implantation. The subsequent annealing at high temperature (1573 K) results in the partial restoration of the structure of the sample.

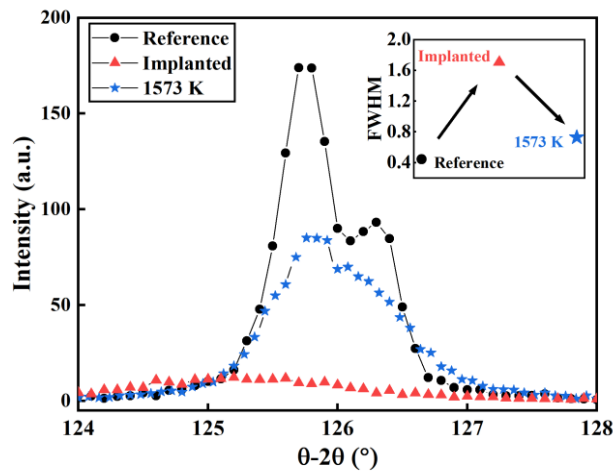

*Figure 3: θ-2θ X-ray diffraction pattern of ScN (333) peak after different steps of investigation of samples (reference, Ar-implanted and annealed at 1573 K).*

The effect of implantation induced damage was studied by quantifying the residual stresses applied to the film. The planes, not parallel to the surface, $33\bar{3}$ were studied (at $\Psi$ of 70.53°) and compared to the 333 reflection; the data are summarized in Table 1. For the reference sample, the inter-planar distances are almost equal between the growth direction and the in-plane direction showing that no (or few) residual stress is present. On the contrary, implanted, and annealed samples have different values of $d_{333}$ and $d_{33\bar{3}}$. After implantation, $d_{333} > d_{33\bar{3}}$ implies that the film undergoes compressive stress due to its expansion after the ion implantation. The same conclusion can be drawn from the stress values evaluated from the "sin² Ψ-method". The stress obtained after implantation is negative, showing the drastic change in the structure and the induced compressive stress. After annealing, the residual stress is found to be positive and the relation $d_{333} < d_{33\bar{3}}$ is observed indicating a strong recovery of the damage and thus a partial restoration of the film structure. However, the value of the residual stress is higher and lead to a compressive stress compared to the one from the reference film, suggesting that annealing has indeed restored the structure and in addition to recombination, a defect evolution has occurred during high temperature annealing.

*Table 1: X-Ray diffraction data for the ScN thin films as-grown, Ar-implanted (5 dpa) and annealed (1573K).*

| Samples | $d_{(333)}$ (Å) | $d_{(33\bar{3})}$ (Å) | σ (GPa) | $a_0$ (Å) |
|---|---|---|---|---|
| Ref. | 0.8656 | 0.8654 | 0.090 ± 0.05 | 4.496 |
| Ar-Implanted | 0.8681 | 0.8670 | -0.320 ± 0.05 | 4.507 |
| Annealed 1573K | 0.8654 | 0.8661 | 0.270 ± 0.05 | 4.497 |

The analysis of the optical properties (refractive index n and extinction coefficient k) of conductive materials gives access to local electrical properties, i.e., the in-grain mobility and carrier density.[24] Figure 4 shows the variation of the optical extinction coefficient *k* versus the incident wavelength. The curve of the reference ScN sample shows different domains: a transparent region from 500 to 900 nm in-between two absorbing regions the inter- and intra-band absorptions. These wavelength ranges are in good agreement with those already measured and calculated using density functional calculations onto single crystal ScN on MgO (001).[25,26] Starting at 900 nm, the curve is representative of the Drude's model suggesting a behavior-like metallic character of ScN. By using an effective electron mass of 0.40$m_0$ the mobility inside the grain at RT has been determined to be $\mu_{grain}^{ref}(RT) \sim 35$ cm²V⁻¹s⁻¹ and the carrier concentration of about 2.1x10²¹ cm⁻³.[25] After implantation, the *k*-curve is drastically altered in agreement with XRD (figure 3). The transparent domain has disappeared, and the electrical nature of the film is nearly lost due to the as-introduced defects preventing any Drude analysis and thus any determination of the electrical properties. The subsequent annealing at 1573 K leads to the recovery of the optical constants and the sample tends to recover its initial color. The carrier concentration is found to be fully restored in contrast to the mobility which is estimated to be $\mu_{grain}^{1573K}(RT) \sim 28\ cm^2V^{-1}s^{-1}$. The optical measurements confirmed that some defects are still present within the ScN grains even after annealing.

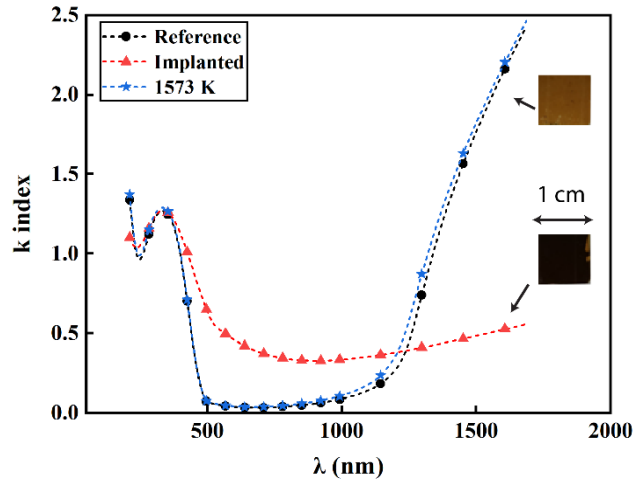

*Figure 4: Extinction coefficient k of ScN film at different steps of the process: reference, Ar-implantation and annealed at 1573K. The optical appearance of the film is presented on the right side.*

Figure 5 shows the TEM analysis of the ScN reference sample along with the implanted and the implanted/annealed one. Figure 5(a) is an overview cross-section TEM image of typical ScN film. Columnar domains are visible highlighting the highly textured polycrystalline character of the film. A thickness of 240 nm can be estimated from the image, in good agreement with the X-Ray Reflectometry (XRR) measurements. After implantation, the implanted zone is clearly visible, figure 5(b). This zone comprises black contrast dots uniformly distributed up to a depth of about 300 nm, in agreement with the SRIM simulations (superimposed in the figure). These small dots cannot be resolved individually and suggest the formation of clusters of defects (interstitials). No other type of defects is observed. EDX-STEM mapping shows the presence of dispersed argon in the film and beyond into the substrate as expected from SRIMS simulations (Fig. 1).

The TEM image, Fig. 5(c), attests to a change of the microstructure happening after annealing at high temperature 1573 K on an ion implanted sample. Firstly, there is the formation of large argon-filled cavities (bubbles evidenced by EDS data) as well as dislocations in the implanted area of the substrate (inset *(i)*). Most bubbles are faceted along basal planes; the others, larger ones, are rather spherical. These observations are similar to those obtained after helium implantation performed on Al$_2$O$_3$ single crystal.[27] In the ScN film the microstructure looks different. There are less black contrast dots than before annealing revealing the columnar structure of the film is still preserved despite all the process. Bubbles in the ScN film are also observed, but they are heterogeneously distributed. Most appear spherical and small, a few nanometers in size (inset *(iii)*), while some defects with a more elongated shape are highly dispersed and can reach 30-40 nm (inset *(ii)*). It would also appear that many bubbles are present at the film/substrate interface resulting in an almost bubble-free zone of about 50nm inside the film, see the inset *(i)* in figure 5(d).

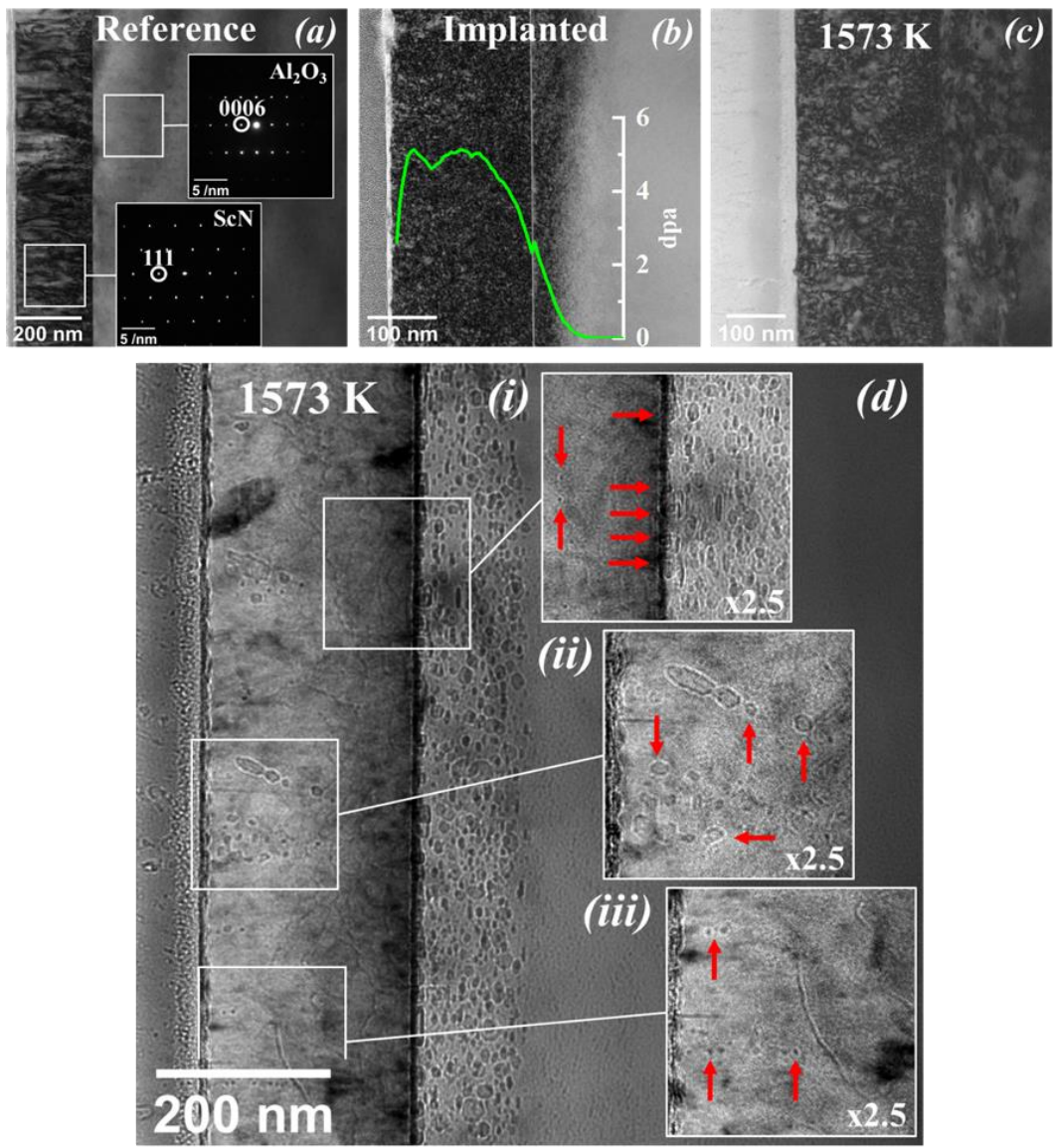

*Figure 5: Bright Field (BF) cross-sectional TEM micrographs of ScN film on Al$_2$O$_3$ (0001) substrate for the reference (a), Ar-implanted (b) and annealed sample at 1573K (c). (d) Detailed BF-image of the annealed sample: tilted with α = -16° and β =0.83° and overfocused. The insets highlight the interface region (i), the large (ii) and tiny bubbles (iii). Bubbles are marked with arrows.*

## 3.2 - PHYSICAL CHARACTERIZATION

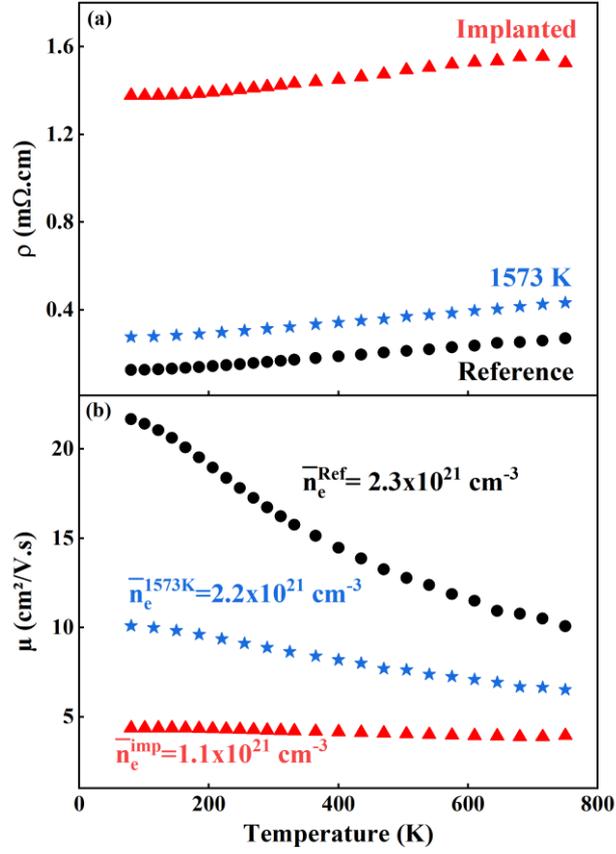

*Figure 6 : Resistivity (a) and mobility (b) as function of temperature of the reference, Ar-implanted and annealed ScN thin films.*

The electrical resistivity measurements are plotted in figure 6(a) in the operating temperature range 80-750 K. All curves show a similar appearance characteristic of metallic-like behavior. With increasing temperature, the electrical resistivity exhibits a constant value from 80 K up to 150 K approximately, and then a linear positive rise. The absence of any thermally activated transport at low temperature suggests an electrical resistivity controlled by impurities and the as-grown crystal defects. In the linear region ($T_0 > 150$ K) the resistivity can be given by the following relation:

$$\rho(T) = \rho_R + \alpha(T - T_0) + \rho_D \qquad (2)$$

with $\rho_R$ the residual resistivity and $\rho_D$ the implantation-induced resistivity ($\rho_D^{ref} = 0$). As seen the slope of curves α~2.2x10$^{-7}$ ± 0.1 Ω.cm.K$^{-1}$ is found to be constant at any stage of the process meaning that the interaction electron-phonon is not modified by the implantation-induced defects and their evolution upon annealing at 1573 K: only $\rho_D$ changes. For the 5 dpa-Ar-implantation $\rho_D^{implanted} = 1.25 \times 10^{-3}$ Ω.cm is found to be temperature independent in all the investigated temperature range, no recovery of damage or defects recombination occurs during the electrical measurements. This suggests that all the defects produced by the Ar implantation are stable up to at least 750 K in contrary to what was observed for Mg dopant implantation in ScN.[18] An annealing at 1573 K restored the electrical resistivity to comparable values as the reference film, the value of $\rho_D^{impl}$ is reduced by about 90%.

Figure 6(b) shows the Hall mobility curves with temperature that helps to understand the carrier scattering mechanism and reports the average of carrier concentrations. According to the Matthiessen's rule, the total mobility can be written as:

$$\frac{1}{\mu(T)} = \sum_i \frac{1}{\mu_i} = \frac{1}{\mu_{lat}(T)} + \frac{1}{\mu_R} + \frac{1}{\mu_D} \qquad (3)$$

where $\mu_{lat}$ is the lattice mobility, $\mu_R$ the residual mobility and $\mu_D$ the implantation-induced defect mobility. For the reference ScN, $\mu_D^{ref}$ is taken as ∞. The Hall mobility curve for the reference sample decreases smoothly with temperature, but more slowly than expected for acoustical lattice scattering for which the relationship is $\mu_{lat}(T) \sim T^{-1.5}$ (or $T^{-1.29}$ as reported in MBE-ScN[28]). It can be well fitted by adding the carrier scattering by residual impurities (as-grown defects and grain boundaries) taken as $\mu_R \sim 23$ cm$^2$V$^{-1}$s$^{-1}$ for T > 150 K. The Hall electron mobility strongly reduced by the argon implantation seems to be constant with temperature (80-750 K), $\mu^{implanted}(T) = $ 4-5 cm$^2$V$^{-1}$s$^{-1}$, with a major contribution of defect's mobility estimated at $\mu_D^{implanted} \sim 5$ cm$^2$V$^{-1}$s$^{-1}$ by using Eq.2. Annealing did not fully restore the total mobility showing the partial recovery of defects and leading to an increase of $\mu_D^{1573K} \sim 18$ cm$^2$V$^{-1}$s$^{-1}$ (T~300K).

      The carrier concentrations of samples are found to be high and independent of temperature which is a typical trait of degenerate semiconductors. The Ar-implantation in this highly damaged regime reduced by a factor two the free carrier concentration showing that the

as-introduced defects act like traps for electrons. In contrast, the carrier concentration was almost recovered during annealing (carrier detrapping). However, defects are still present after annealing and affect the charge carrier mobility while being no longer electrically active. These electrical measurements are in good agreement with the optical characterizations, namely a fully recovery of the carrier concentration after annealing and a partial recovery of the carrier mobility.

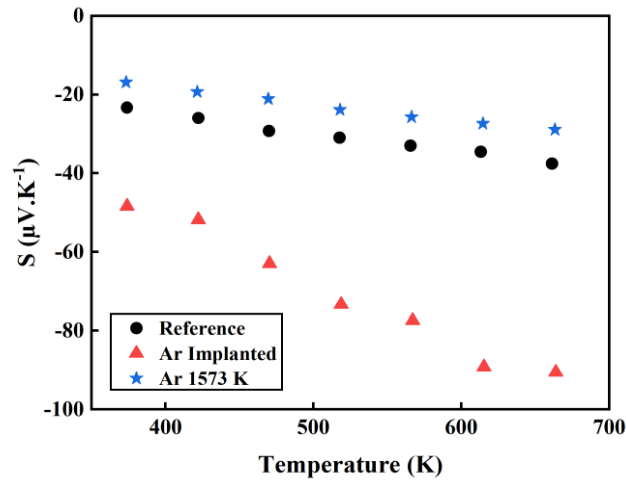

*Figure 7: Temperature dependence of the Seebeck coefficient (S) from 370 to 680K of the reference, the Ar-implanted (5 dpa) and annealed (1573K) samples.*

Seebeck coefficients measured in the temperature range 370-680 K are displayed in figure 7. $|S|$ increases linearly with the measuring temperature for all the samples regardless of process step. The linear dependence with temperature suggests a constant carrier concentration in agreement with Hall-effect measurements (figure 6(b)) At 600 K the value of $|S|$ for the reference ScN is 35 µV.K$^{-1}$ being close to the value previously reported for samples produced under similar conditions but lower than the one reported for ScN growth on MgO substrate by using MBE.[15,18] This low value may be due to the large contamination from impurities acting as dopants resulting in the large carrier concentration measured. After implantation, the Seebeck coefficient absolute values in the high damage regime increases from 30 to 85 µV.K$^{-1}$ at 600 K. As seen the implantation process also increases the slope of the curve by a factor of close to three. After annealing at 1573 K, the absolute value of the Seebeck coefficient is found to be slightly lower than reference value, ~ 28 µV.K$^{-1}$ at 600 K, with, however, the slope back to the same value. The Seebeck coefficient must therefore be analyzed considering both the number of carriers and the presence of defects that change the Density of States (DOS).

*Table 2 : Identified thermal conductivity, contact resistance and heat capacity for reference, Ar-implanted and annealed ScN thin films*

| Samples | Ref. | Ar-Implanted | Annealed 1573K |
|---|---|---|---|
| $k_{total}$ ($W.m^{-1}.K^{-1}$) @ 300 K | 12.5 | 3 | 8.5 |
| $R_{contact}$ ($nK.m^2.W^{-1}$) @ 300 K | 18 | 24 | 14 |
| $C_{TH}$ ($MJ.K^{-1}·m^{-3}$) @ 300 K | 4.07 | 4.03 | 4.1 |

Table 2 reports the values of the thermal conductivity of ScN measured at 300 K, the heat capacity of the ScN thin film and the interface thermal resistance between the top aluminum film and the layer of interest. The thermal conductivity value of the reference sample is 12.5 W.m$^{-1}$K$^{-1}$, in good agreement with the values reported earlier.[14,18] After Ar-implantation, the sample exhibits a large drop of thermal conductivity down to 3 W.m$^{-1}$K$^{-1}$. As previously mentioned, annealing process afterwards restored partially the structure and the defects, leading a thermal conductivity towards its original value (around 8.5 W.m$^{-1}$K$^{-1}$) and highlighting again the presence of structural defects even after high temperature annealing. The heat capacity is not affected by the entire process. The contact thermal resistance, increase slightly after implantation and it recovers a value comparable to the reference sample after annealing.

## 4 – DISCUSSION

The large decrease of thermal conductivity in implanted ScN thin films either with noble gas (Ar in this study, see table 2), with dopants (Mg in a previous study) or by non-electrically active element (Li$^+$) may be explained by the as-introduced defects which reduce the mean phonons free path, increasing thus the level of scattering.[18,19] The increase of thermal conductivity occurring during the subsequent high-temperature annealing suggests a partial recovery of defects. However, the Seebeck curves and electrical characterizations provide a more complete picture. The implantation defects introduce deep acceptors level in the band gap then reduce the concentration of free carriers. Moreover, this should modify locally the electronic DOS. For degenerate semiconductors, the Seebeck coefficient is dependent on the effective mass of carriers at the Fermi surface.[7] Besides, calculations showed that vacancies introduce asymmetrical peak close to the Fermi level in the electronic DOS of ScN, resulting in an enhancement of the Seebeck coefficient.[29,30] As a result, the trapping of carriers and the modification of the DOS caused by the implantation process led to an increase of both Seebeck value and slope (x3 in the implanted sample) of the curve *S(T)* (figure 7), in the whole investigated temperature range. This increase in slope is also observed on the previous study when implanting Mg-dopants in ScN.[18] A higher *S(T)* slope was reported when measuring the thermoelectric properties up to the temperature at which the defects recombination starts to be active (any zero-dimensional defects such as the Frenkel pairs); i.e., at 450 K.[18] This triggering of defect recombination results also in a progressive decrease of electrical resistivity during the measurement. This behavior was reported regardless of the concentration of implanted Mg.

In the present paper, no change in slope is observed when implanting argon at high fluence (figure 6); the Ar-atoms therefore operate as point defect stabilizers preventing any defect recombination (or damage recovery) at least up to a temperature of 750 K (see the electrical resistivity and Seebeck curves Fig. 2 and 3, no modifications occur during temperature range). Noble gas (NG) atoms are known to behave singularly when implanted in materials.[20] Because of their low solubility in materials, they tend to aggregate resulting in the formation of NG-extended defect such as cavities or highly pressurized bubbles in fluid or in solid form depending on implantations conditions.[31] Electronic structure calculations in SiC and Si showed that the trapping of argon (as other heavy gases) by mono- or divacancy is energetically favorable.[32,33] Implanted argon atoms in ScN are thus expected to be trapped by the supersaturation of vacancies introduced by the collision cascades. Thus, during the implantation process, many interstitials recombine with vacancies (dynamical annealing) or combine with other to form interstitial clusters that appears as black spots damage in TEM (figure 5(a)). All the remaining vacancies trap the argon atoms to form argon-vacancies complexes, $Ar_nV_m$ with $m>n$. Up to 750 K neither the interstitials clusters nor the $Ar_nV_m$ complexes have sufficient energy to dissociate and no change in the slopes of the Seebeck and resistivity curves is reported with the measuring temperature (figures 6 and 7).

All these defects contribute predominantly to scatter the free carriers and therefore to reduce drastically the electrical mobility. After annealing, the recovery of the Seebeck slope shows the removing of the DOS changes caused by the implantation. The values of |*S*| are however slightly lower than in the reference sample, highlighting the change in type of defects. Moreover, this evolution is coupled with a change in the stress state of the film from compressive to tensile.

The different as-introduced defects, have an influence on the mobility of free carriers by reducing their free mean path. The black dots damage observed after implantation suggests that the primary knock-on-atoms go on to generate collision cascades in which Frenkel pairs are formed. Probably due to the low migration energies if many interstitials recombine with vacancies others combine to form clusters appearing as black spot damage on the TEM image, figure 5(b). According to the temperature of subsequent annealing such clusters have sufficient energy to dissociate giving rise to a lower density of larger black-spots as seen in figure 5(c) when compared to figure 5(b). The continuous recovery of the resistivity with annealing suggests a size dependence of the dissociation energies of the clusters. The role of Ar-gas leads to the stabilization of vacancy-type defects as cavities (bubbles or voids) are observed before annealing. Upon annealing, the TEM observations suggest that some of the ($Ar_nV_m$) complexes dissociate/migrate to other complexes until they form visible bubbles. However, their formation and growth appear to be the result of a combination of several factors including stresses evolution, film structure (as the presence of grains) and defect mobility resulting in a rather heterogenous distribution of bubbles in the film. As an example, the growth of bubbles in SiC is found to be enhanced on grain boundaries.[34] In ScN, by applying Matthiessen's rules to optical and Hall mobilities at RT, the mobility due to grain boundaries is found to be reduced after implantation and annealing from $\mu_{GB}^{ref}(RT) \sim 35 \text{ cm}^2\text{V}^{-1}\text{s}^{-1}$ to $\mu_{GB}^{1573K} \sim 13 \text{ cm}^2\text{V}^{-1}\text{s}^{-1}$. This shows that the grain boundary scattering is not neglectable for carrier mobility and that an accumulation of defects also occurs on the grain boundaries during the 1573K annealing. Similarly, the interface (14% lattice mismatch) appears to act as a sink for bubbles formation. TEM observations are still in progress to obtain a clearer picture. All these defects will therefore influence the transport properties of ScN.

Combining the Seebeck and electrical conductivity values the power factor at 600 K for the reference sample is found to be about 5x10$^{-4}$ Wm$^{-1}$K$^2$. This value is relatively low compared to the previous studies on as-grown and Mg-implanted ScN due to the quality of the film;[13,14,18] i.e., the unwanted dopant impurities introduced during the film synthesis as the amount of oxygen or fluorine. The quality of the ScN film is a critical factor in improving its thermoelectric properties. While implantation damage leads to a strong increase in the value of the Seebeck coefficient, it also strongly reduces the electrical conductivity, which in turn has no or few effects on the calculated value of the power factor, ~5x10$^{-4}$ Wm$^{-1}$K$^2$. On the contrary, the damage induces a strong reduction in thermal conductivity which is beneficial for improving the thermoelectric figure of merit of ScN, and these defects are stable to at least 750 K. Post implantation annealing results in an evolution of defects leading to a de-trapping of free carriers but these defects (interstitials clusters, bubbles, vacancies-gas complexes: 3D-defects) still affect the mobility and also reduce the thermal conductivity which was the purpose of this study. Then, knowing that the types of defects are strongly dependent on the implantation and post-processing conditions, these must be optimized to find the best balance between all the interrelated parameters ($S$, $\rho$ and $k_{total}$).

## 5 - CONCLUSIONS

Argon implantation at room temperature in a high regime of damage (5-6 dpa) was carried out on ScN thin film to introduce defects in order to reduce the lattice thermal conductivity. All implantation defects created are stable up to a minimum operating temperature of 750K: the use of argon therefore stabilizes implantation induced-defects. These defects act as carrier traps, reduce their mobility and also have a strong effect on the thermoelectric properties of ScN. In particular, the thermal conductivity is found to be reduced by a factor of four while keeping the PF constant. Post implantation annealing at high temperature restores the crystallinity of the ScN structure and the number of free carriers but also leads to the formation of nanosized 3D-defects which both affect carrier mobility and phonon scattering. However, these defects have a detrimental effect on the power factor showing that the thermoelectric properties are strongly dependent on the size, morphology, and type of defects.

Thus, this study shows that the controlled introduction of defects, or the defect engineering (via noble gas implantation for thin films), can be used as a strategy to introduce additional phonon scattering centers that is beneficial for the development of TE-materials.

**ACKNOWLEDGMENTS**

This work was supported by the French government program "Investissements d'Avenir" (EUR INTREE, reference ANR-18-EURE-0010 and LABEX INTERACTIFS, reference ANR-11-LABEX-0017-01). This work was supported by the ANR Project SPIDER-man ANR-18-CE42-0006. This work has also been partially supported by « Nouvelle Aquitaine » Region and by European Structural and Investment Funds (ERDF reference: P-2016-BAFE-94/95; CHARTS 2019-1R1M04). The authors also acknowledge funding from the Swedish Research Council (VR) under Project No. 2021-03826, the Knut and Alice Wallenberg Foundation through the Wallenberg Academy Fellows program (grant no. KAW 2020.0196), the Swedish Government Strategic Research Area in Materials Science on Functional Materials at Linköping University (Faculty Grant SFO-Mat-LiU No. 2009 00971), and the Swedish Energy Agency under project 46519-1.